\documentclass[twoside,reqno]{article}
\usepackage[tbtags]{amsmath}
\usepackage{amsfonts}
\usepackage[mathcal]{eucal}
\usepackage[pdfstartview=FitH,pdfnewwindow=true,pdftitle={Canonical formalism for quasi-classical particle 'Zitterbewegung'},pdfdisplaydoctitle=true,colorlinks,pagebackref,bookmarksopen,breaklinks]{hyperref}
\usepackage{amssymb}
\newcommand\so{{\scriptscriptstyle 0}}
\newcommand\sv{\mathsf v}
\newcommand\bv{\boldsymbol{\mathsf v}}

\newcommand\bu{\boldsymbol{\mathsf u}}
\newcommand\buu{\boldsymbol{\dot{\mathsf u}}}
\newcommand\buuu{\boldsymbol{\ddot{\mathsf u}}}
\newcommand\uo{u_{\scriptscriptstyle 0}}
\newcommand\bwp{\pmb{\boldsymbol\wp}}

\newcommand\bp{\boldsymbol{\mathsf p}}
\newcommand\bcd{\boldsymbol\cdot}
\newcommand\cc{\!\circ\!}

\newcommand\bx{\boldsymbol{\mathsf x}}
\newcommand\2{{^{{\boldsymbol 2}}}}
\begin{document}
\title{\rule{0pt}{0pt}\\[-5.5cm]Canonical formalism for quasi-classical particle `Zitterbewegung' in Ostrohrads{\kern-1pt'\kern.3pt}kyj mechanics}
\author{\href{http://www.iapmm.lviv.ua/12/eng/files/st_files/matsyuk.htm}{Roman Matsyuk}\\Institute for Applied Problems in Mechanics and
Mathematics,
\\ 15~Dudayev~St., 290005 L\kern-1pt'viv, Ukraine
\\
{Email:~\texttt{romko.b.m@gmail.com, matsyuk@lms.lviv.ua}}
}
{
\renewcommand{\thefootnote}{}
 \footnotetext{MS 2000
classification: 70H50}
\footnotetext{Research supported by grant GACR 201/03/0512 of the Czech
Grant Agency}
\date{\href{http://www.karlin.mff.cuni.cz/~dgac/proceedings/Matsyuk.ps}{Proc. DGA 9 2004. Published 2005.} Corrected June 22, 2014.}
\maketitle
}
\begin{abstract}
The homogeneous canonical formalism of Rund is applied to the
se\-cond-order Lagrangian model of the self-interacting particle of
Bopp. The quasi-classical free spinning particle of Mathisson
appears then as a constrained subsystem of the previous.
Differential-geometrical mechanisms offered are formulated in a fairy
general manner, although revealed here in a particular example of
physical meaning.

\end{abstract}

\section {A brief overview of Rund prescription}
In his book on the Hamilton-Jacobi theory~\cite{matsyuk:Rund} Hanno
Rund proposed a pa\-ra\-me\-ter-homogeneous formulation of the
variational problem with higher derivatives. An appropriate
prescription was given there for the transition to Hamiltonian formalism
which is best suited to the needs of relativistic mechanics, and,
 generally speaking, especially convenient in all those cases, where an
invariance with respect to some transformation group of all variables
(dependent and independent ones) is imposed by the very prerequisites
of the theory.

Let
\begin{equation}\label{matsyuk:pr}
pr\!:T^{r}M\setminus\{0\} \to C^{r}(1,M)
\end{equation}
denote the quotient projection of the manifold of non zero
Ehresmann velocities to the manifold of contact elements of $r$-th
order with respect to the action the (local) reparametrization
group $\mathrm{Gl^{r}}(1,\mathbb R)$ on $T^{r}M$. Every time a
Lagrange function $\mathcal L\!: T^{r}M \mapsto \mathbb R$
satisfies the so-called Zermelo conditions, it defines a
parameter-invariant variational problem on $T^{r}M$. Every such
problem passes to the above mentioned quotient and defines certain
sheaf of equivalent semi-basic $1$-forms (or Lagrangian densities)
on the fibred manifold $C^{r}(1,M)$ over $M$. The general setting
for this mechanism was discussed in details
in~\cite{matsyuk:DGA8}, the usage of sheaf theory concepts was
justified by Paul Dedecker~\cite{matsyuk:Dedecker}. In present
contribution we shall limit ourselves to the case of order $2$
($r=2$) variational problem and, moreover, shall work in local
coordinate representation to touch with the physical model as
announced. The convenient commonly accepted coordinates in the
manifolds introduced as far read $x^{\alpha}$, $u^{\alpha}$, $\dot
u^{\alpha}$, $\ddot u^{\alpha}$, $\dddot u^{\alpha}$ for $T^{4}M$
and $x^{\so}$, $x^{i}$, $\sv^{i}$, $\sv'^{i}$, $\sv''^{i}$,
$\sv'''^{i}$ for $C^{4}(1,M)$. As soon as in our application $M$
becomes the space-time of special relativity with the diagonal
metrics $(1, -1, -1, -1)$, we shall put to use vector notations of
the pattern $u=(u_{\so}, \bu)$, $u\cdot u=u_{\so}^{2}+\bu\2$,
$\bu\2=\bu\bcd\bu=u_{\alpha}u^{\alpha}$.

Let $\mathcal L (x, u, \dot u)$ be a Lagrange function on $T^{2}M$
that satisfies Zermelo conditions:
\begin{equation}\label{matsyuk:Zermelo}
\begin{gathered}
u^\alpha \dfrac{\partial \mathcal L}{\partial \dot u^\alpha} \equiv 0 \\
u^\alpha \dfrac{\partial \mathcal L}{\partial u^\alpha} +
2\,\dot u^\alpha \dfrac{\partial \mathcal L}{\partial \dot u^\alpha} \equiv
\mathcal L \,.
\end{gathered}
\end{equation}
As common, let us introduce the Legendre
transformation $Le\!:(x, u, \dot u, \ddot u) \mapsto (x, u, \wp, \wp')$,
\begin{equation}\label{matsyuk:cal p}
\begin{gathered}
\wp'=\dfrac{\partial \mathcal L}{\partial \dot u} \\
\wp=\dfrac{\partial \mathcal L}{\partial u} - \mathcal D_\tau \wp'\,,
\end{gathered}
\end{equation}
where
\begin{equation}\label{matsyuk:Dtau}
\mathcal D_\tau = u \frac{\partial }{\mathstrut\partial x} +
\dot u \frac{\partial }{\mathstrut\partial u} +
\ddot u \frac{\partial }{\mathstrut\partial \dot u}
\end{equation}
denotes the operator of total derivative. We also mention that in forthcoming
application nothing depends on space-time variables $x$, since everything obeys
the pseudo-Euclidean symmetry.

It may be seen that the Zermelo conditions, if fulfilled, are now equivalent
to the following:
\begin{subequations}\label{matsyuk:Z}
\renewcommand{\theequation}{\theparentequation .\arabic{equation}}
\begin{gather}
u^\alpha\wp'_\alpha \equiv 0 \label{matsyuk:Z1}\\
u^\alpha\wp_\alpha + \dot u^\alpha\wp'_\alpha \equiv \mathcal L\,.
\label{matsyuk:Z2}
\end{gather}
\end{subequations}

According to H.~Rund, we assume that there exists a $C^2$ function $\mathcal
H$ of the four variables $(x, u, \wp, \wp')$ which is not trivially constant along each of the last
two  variables, which is nevertheless constant along the Legendre
transformation, and we chose that constant to
be equal to $1$ without any essential loss of generality:
\begin{equation}\label{matsyuk:H=1}
\mathcal H\circ Le\equiv 1\,.
\end{equation}

As proved in~\cite{matsyuk:Rund} (see also~\cite{matsyuk:Grasser}),
under the assumption that
\begin{equation}\label{matsyuk:rank}
\text{rank}
\left\|\frac{\partial^2 \mathcal L}{\partial \dot u^\alpha \partial \dot
u^\beta}\right\| = \dim M-1\,,
\end{equation}
there exist proportionality factors $\lambda$
and $\mu$, in general dependent on $x, u, \dot u, \ddot u$, such that the
following {\it canonical system} of differential equations of the first order
with respect to the variables $x, u, \wp, \wp'$ is satisfied along all the
extremals of the variational problem with the Lagrange function $\mathcal L$:
\begin{subequations}\label{matsyuk:H}
\renewcommand{\theequation}{\theparentequation .\arabic{equation}}
\begin{align}
\frac{dx}{d\tau}&=\lambda\,\frac{\partial
\mathcal H}{\partial \wp} \label{matsyuk:Hx}\\
\frac{du}{d\tau}&=\lambda\,\frac{\partial
\mathcal H}{\partial \wp'} + \mu u \label{matsyuk:Hu}\\
\frac{d\wp}{d\tau}&=-\lambda\,\frac{\partial
\mathcal H}{\partial x} \label{matsyuk:Hp}\\
\frac{d\wp'}{d\tau}&=-\lambda\,\frac{\partial
\mathcal H }{\partial u} - \mu \wp' \,. \label{matsyuk:Hp'}
\end{align}
\end{subequations}

Now the evolution of an arbitrary function $f$ of the phase space
variables $x$, $u$, $\wp$, $\wp'$ is given by the famous Poisson bracket
\[
\big\{f,\mathcal H\big\}\overset{\mathrm{def}}= \dfrac{\partial f}{\partial x^\alpha}
        \dfrac{\partial \mathcal H}{\partial \wp_\alpha} +
 \dfrac{\partial f}{\partial u^\alpha}
        \dfrac{\partial \mathcal H}{\partial \wp'_\alpha} -
 \dfrac{\partial f}{\partial \wp_\alpha}
        \dfrac{\partial \mathcal H}{\partial x^\alpha} -
 \dfrac{\partial f}{\partial \wp'_\alpha}
        \dfrac{\partial \mathcal H}{\partial u^\alpha}
\]
as follows~\cite{matsyuk:Grasser}
\begin{equation}\label{matsyuk:Poisson}
\dfrac{df}{d\tau} = \lambda \big\{f,\mathcal H \big\}
+ \mu \left[u^\alpha\dfrac{\partial f}{\partial u^\alpha} -
\wp'_\alpha \dfrac{\partial f}{\partial \wp'_\alpha}\right]\,.
\end{equation}

\section{How to obtain the $\mathcal H$}

The scope of possible functions {$\mathcal H$} who
satisfy~(\ref{matsyuk:H=1}) is rather large. But, since every
pa\-ra\-me\-ter-independent variational problem, posed on $T^rM$, generates
a corresponding formulation on $C^r(1,M)$, and vice versa, one may
effectively try a pull-back of the Hamiltonian formulation of
the problem on $C^r(1,M)$ to $T^rM$.

Let a variational problem on $\mathbb R\times T^rM$ be given in terms of the
semi-basic (relative to $\mathbb R$) differential $1$-form $\mathcal L \,
d\tau$, where $\mathcal L$ is defined on $T^rM$ solely and satisfies the
Zermelo conditions. And let $L\, dx^{\so}$ be that representative of the
corresponding sheaf of equivalent semi-basic (relative to $M$) differential
$1$-forms on the fibred manifold $C^r(1,M)$, who in the above described coordinates
is given by the following relation,
\begin{equation*}
\mathcal L\, d\tau  -  (L\cc pr)\, dx^{\so} = {} - (L\circ pr)\,\vartheta \,,
\end{equation*}
where
\begin{equation}\label{matsyuk:theta}
\vartheta = dx^{\so} - u^{\so}d\tau
\end{equation}
is one of the contact forms on $J^1(\mathbb R,M)\approx \mathbb R\times TM$.
Hence\nopagebreak
\begin{equation}\label{matsyuk:cal L}
\mathcal L=u^{\so}\,L\circ pr.
\end{equation}
The canonical momenta are being introduced here as usual:
\begin{equation}\label{matsyuk:bp}
\begin{gathered}
\bp'=\dfrac{\partial L}{\partial \bv'} \\
\bp=\dfrac{\partial L}{\partial \bv} - D_t \bp'\,,
\end{gathered}
\end{equation}
where
\begin{equation}\label{matsyuk:Dt}
D_t = \sv^{i} \frac{\partial }{\mathstrut\partial x^i} +
\sv'^i \frac{\partial }{\mathstrut\partial \sv^i} +
\sv''^i \frac{\partial }{\mathstrut\partial \sv'^i}
\end{equation}
denotes the operator of total derivative with respect to $x^{\so}$.

The  correspondence between
the operators (\ref{matsyuk:Dtau}) and (\ref{matsyuk:Dt}) of total
derivatives on relevant jet spaces, $J^2(\mathbb R, M)$ and (locally)
$J^2(\mathbb R, \mathbb R^{\,\mathrm {dim}M-1})$
seems evident, as in fact it is: whenever $\mathsf f$ is a local function on
$C^2(1,M)$, then
\begin{equation}\label{matsyuk:Dtau=Dt}
\mathcal D_\tau(\mathsf f\circ pr)=u^{\so}\,D_t \mathsf f\circ pr
\end{equation}
It would, however, be
of some instructive good to obtain (\ref{matsyuk:Dtau=Dt}) by direct
differentiation of the projection (\ref{matsyuk:pr}), which, in the
3\textsuperscript d order, reads in our coordinates:
\begin{equation}\label{matsyuk:u=v}
\begin{gathered}
\bv\circ pr=\dfrac\bu {u_{\so}} \\
\bv'\circ pr=\dfrac{\buu}{u_{\so}^2} - \dfrac{\dot
u_{\so}}{u_{\so}^3}\,\bu\\
\bv''\circ pr=\dfrac{\buuu}{u_{\so}^3} -
3\,\dfrac{\dot u_{\so}}{u_{\so}^4}\,\buu + 3\left(\dfrac{\dot
u_{\so}^2}{u_{\so}^5} - \dfrac{\ddot u_{\so}}{u_{\so}^4}\right)\bu\,.
\end{gathered}
\end{equation}

With relation~(\ref{matsyuk:Dtau=Dt}) in hand, we are ready now
to establish the correspondence between the pair of
momenta $\wp=(\wp_{\so}, \bwp)$ and $\wp'=(\wp'_{\so}, \bwp')$
in~(\ref{matsyuk:cal p}), calculated for the Lagrange function
$\cal L$ given by (\ref{matsyuk:cal L}), and the pull-back of the
momenta in~(\ref{matsyuk:bp}): {\allowdisplaybreaks
\begin{subequations}\label{matsyuk:all p=p}
\renewcommand{\theequation}{\theparentequation .\arabic{equation}}
\begin{align}
\label{matsyuk:wp'0=p'0}
\wp'_{\so}&=\uo\dfrac{\partial (L\cc pr)}{\partial \dot u_{\so}}=
        -\dfrac1{\uo^2}\,\bu\left(\dfrac{\partial L}{\partial \bv'}\cc pr\right)=
        -\dfrac1{\uo^2}\,\bu\,(\bp'\cc pr)\,;\\
\label{matsyuk:bwp'=bp'}
\bwp'&=\uo\dfrac{\partial (L\cc pr)}{\partial \buu}=
        \dfrac1{\uo}\left(\dfrac{\partial L}{\partial \bv'}\cc pr\right)=
        \dfrac1{\uo} (\bp'\cc pr)\,;
\end{align}
\begin{align}
\label{matsyuk:wp0=p0}
\begin{split}
\wp_{\so}&=L\cc pr+\uo\,\dfrac{\partial (L\cc pr)}{\partial u_{\so}}
        -\mathcal D_\tau \wp'_\so \quad \text{by the reason of (\ref{matsyuk:u=v}), (\ref{matsyuk:Dtau=Dt}) and (\ref{matsyuk:wp'0=p'0})}
        \\
         &=L\cc pr
         -\uo\left[
          \dfrac1{\uo^2}\,\bu\left(\frac{\partial L}{\partial \bv}\cc pr\right)
         +\dfrac{2}{\uo^3}\,\buu\left(\frac{\partial L}{\partial \bv'}\cc pr \right)
         -\dfrac{3\dot u_{\so}}{\uo^4} \,\bu\left(\dfrac{\partial L}{\partial\bv' }\cc pr\right)\right] \\
         &\phantom{=L\cc pr\;}-2\dfrac{\dot u_\so}{\uo^3}\,\bu(\bp'\cc pr)+\dfrac{1}{\uo^2}\,\buu (\bp'\cc pr)
         +\dfrac1{\uo}\,\bu(D_t\bp'\circ pr)\\
         &=L\cc pr
          -\dfrac1{\uo}\,\bu\left(\frac{\partial L}{\partial \bv}\cc pr\right)
          +\dfrac{\dot u_\so}{\uo^3}\,\bu(\bp'\cc pr)-\dfrac{1}{\uo^2}\,\buu (\bp'\cc pr)
         +\dfrac1{\uo}\,\bu(D_t\bp'\circ pr)\\
        &=L\cc pr -\bv\bp\circ pr -\bv'\bp'\circ pr\,;
\end{split}
\\[3\jot]
\label{matsyuk:bwp=bp}
\begin{split}
\bwp&=\uo\,\dfrac{\partial (L\cc pr)}{\partial \bu}
        -\mathcal D_\tau \bwp' \quad \text{by the reason of (\ref{matsyuk:u=v}), (\ref{matsyuk:Dtau=Dt}) and (\ref{matsyuk:bwp'=bp'})}
        \\
         &=\uo
         \left[
          \dfrac1{\uo}\,\left(\dfrac{\partial L}{\partial \bv}\cc pr\right)
         -\dfrac{\dot u_{\so}}{\uo^3} \left(\dfrac{\partial L}{\partial\bv' }\cc pr\right)
         \right]
         +\dfrac{\dot u_\so}{\uo^2}\,(\bp'\cc pr)-D_t\bp'\circ pr)\\
         &=\dfrac{\partial L}{\partial \bv}\circ pr - D_t \bp'\circ pr = \bp\circ pr\,.
\end{split}
\end{align}
\end{subequations}
}
From (\ref{matsyuk:wp0=p0}) and (\ref{matsyuk:bwp=bp}) it follows that
{\allowdisplaybreaks
\begin{subequations}\label{matsyuk:wp=p final}
\renewcommand{\theequation}{\theparentequation .\arabic{equation}}
\begin{align}
\wp\, u&=\uo\, L\cc pr - \uo\,\bv'\bp'\circ pr\,,\\
\intertext{whereas from (\ref{matsyuk:wp'0=p'0}) and
(\ref{matsyuk:bwp'=bp'}) in view of (\ref{matsyuk:u=v}) it follows that}
 \wp'\dot u&= \uo\,\bv'\bp'\circ pr\,,
\end{align}
\end{subequations} } and (\ref{matsyuk:Z2}) keeps true immediately.

In our further considerations we choose the approach
of the generalized Hamiltonian theory as
exposed in \cite{matsyuk:Krupkova}. Then, in our coordinates, it is best to
describe the system evolution by the kernel of the differential two
form
\begin{equation}\label{matsyuk:H form}
\omega=-dH\wedge dx^{\so}  + d\bp\wedge d\bx + d\bp'\wedge d\bv\,,
\end{equation}
where the exterior product sign $\wedge$ comprises the contraction
of vector differential forms, if necessary. Now, it is tentative
that on the manifold $\mathbb R\times T^3M$ the evolution of this
same system be described by a differential two form of the same
shape,
\begin{equation}\label{matsyuk:cal H form}
\varOmega=-d\,\mathcal H\wedge d\tau  + d\wp\wedge dx + d\wp'\wedge du\,,
\end{equation}
where the momenta $\wp$ and $\wp'$ due to the Lagrange
function~(\ref{matsyuk:cal L}).

Conventionally one puts
$H=\bp\bv+\bp'\bv'-L$.
Under this assumption it is straightforward to calculate the difference
between (\ref{matsyuk:cal H form}) and (\ref{matsyuk:H form}), taking into
account the relations
(\ref{matsyuk:bwp'=bp'}, \ref{matsyuk:bwp=bp})
and the Zermelo condition
(\ref{matsyuk:Z1}):

\begin{equation}\label{matsyuk:W-w}
\varOmega-pr^*\omega=
d(pr^*H+\wp_\so )\wedge dx^{\so}
-d\,\mathcal H\wedge d\tau\,.
\end{equation}
We wish that this difference be proportional to the contact form
(\ref{matsyuk:theta}), namely,

\begin{equation}\label{matsyuk:20mod}
\varOmega-pr^*\omega=\alpha\wedge\vartheta\,.
\end{equation}

The simplest reasonable way to comply in (\ref{matsyuk:W-w}) with
(\ref{matsyuk:20mod}) is to put

\begin{equation}\label{matsyuk:dcalH}
d\,\mathcal H=u^{\so}d(pr^*H+\wp_{\so})
\end{equation}
and
\begin{equation}\label{matsyuk:psi}
\mathcal H=\uo pr^*H+\varPsi\,.
\end{equation}

Now proceed to determine this deviating function $\varPsi$.
From (\ref{matsyuk:psi}) we have:

\begin{equation}\label{matsyuk:pr*dH}
pr^*dH=\dfrac{d\,\mathcal H}{\uo}
+ \left(\varPsi-\mathcal H \right)\dfrac{d\uo}{\uo^2}-\dfrac{d\,\varPsi}{\uo}\,.
\end{equation}
It suffices now to substitute (\ref{matsyuk:pr*dH}) into (\ref{matsyuk:dcalH})
to obtain the relation

\[
\dfrac{\mathcal H-\varPsi}{\uo}\,d\uo-\uo\,d\wp_\so ={}-d\,\varPsi\,,
\]
from where it becomes clear that
\[\begin{cases}
\varPsi=\uo\wp_\so+c\\\mathcal H=c
\end{cases}\]
and also, by the reason of (\ref{matsyuk:H=1}), $c=1$.

Hence
\begin{equation}\label{matsyuk:cal H}
\mathcal H=\uo pr^*H+\uo\wp_\so +1
\end{equation}

\section{{\itshape Zitterbewegung} of quasiclassical relativistic\\ particle}

As far back as 1946 Fritz Bopp developed a second-order Lagrange
function for the description of classical particle motion from the
second step approximation with respect to the parameter of retard
interaction~\cite{matsyuk:Bopp}. It seems prominent that the Bopp
Lagrangian may be cast into a simple shape in terms of the first
curvature of the particle's world line,
\begin{equation*}\label{matsyuk:k in u}
k=\dfrac{\|\dot u\wedge u\|}{\|u\|^3}\,,
\end{equation*}
as follows:
\begin{equation}\label{matsyuk:Bopp}
\mathcal L\overset{\mathrm{def}}= a \mathcal L_r + A \mathcal L_e
        = \frac a 2 \|u\| k^2+\frac A 2 \|u\|\,,
\end{equation}
where we assume $a\not=0$ to confine with~(\ref{matsyuk:rank}).
This Lagrange function satisfies the Zermelo conditions~(\ref{matsyuk:Zermelo}).
The first addend in~(\ref{matsyuk:Bopp}), $\mathcal L_r$,  turns out to be of the type considered
by H.~Rund in~\cite{matsyuk:Rund} (see also~\cite{matsyuk:Grasser}).
The second addend, $\mathcal L_e$, is the free particle Lagrange function.
According to (\ref{matsyuk:cal L}), the corresponding local Lagrange density
on $C^2(1,M)$ may be expressed in coordinates $x^{\so}$, $\bv$ and
$\bv'$:
\begin{multline}\label{matsyuk:L}
L\,dx^{\so}\overset{\mathrm{def}}= a L_r dx^{\so}+A L_e dx^{\so}\\
        = \frac{a}{2}\sqrt{(1+\bv\2)}
        \left(\frac{\bv'\2}{(1+\bv\2)^2}-\frac{(\bv\bcd\bv')^2}{(1+\bv\2)^3}\right)dx^{\so}
        +\frac{A}{2}\sqrt{(1+\bv\2)}dx^{\so}\,.
\end{multline}
The momenta (\ref{matsyuk:bp}) for this Lagrangian read:
\[
\bp'_r=\dfrac{\bv'}{(1+\bv\2)^{3/2}}-\dfrac{\bv\bcd\bv'}{(1+\bv\2)^{5/2}}\,\bv \\
\]
\begin{multline*}
\bp_r=-\dfrac{\bv''}{(1+\bv\2)^{3/2}}+3\,\dfrac{\bv\bcd\bv'}{(1+\bv\2)^{5/2}}\,\bv' \\
+\dfrac{\bv\bcd\bv''}{(1+\bv\2)^{5/2}}\,\bv
-\dfrac{1}{2}\dfrac{\bv'\2}{(1+\bv\2)^{5/2}}\,\bv
-\dfrac{5}{2}\dfrac{(\bv\bcd\bv')^2}{(1+\bv\2)^{7/2}}\,\bv\,.
\end{multline*}
We introduce the standard
Hamilton function
\begin{multline}\label{matsyuk:serif H}
H=\bp\bv+\bp'\bv'-L
        \\ \overset{\mathrm {def}}=
        aH_r+AH_e=a\bp_r\bv+a\bp'_r\bv'-aL_r+A\bp_e\bv-AL_e\,,
\end{multline}
because $\bp'_e=0$. It is necessary to exclude the variable $\bv'$
in~(\ref{matsyuk:serif H}). We calculate:
\[
\begin{cases}
\bp'_r\bv'=2L_r \\
\bp'_r\2+(\bp'_r\bv)^2=2\,\dfrac{L_r}{(1+\bv\2)^{3/2}}\,,
\end{cases}
\]
and finally the Hamilton function reads
\begin{equation}\label{matsyuk:H final}
H=\bp\bv+\frac 1{2a}\left(1+\bv\2\right)^{3/2}\left(\bp'\2+(\bp'\bv)^2\right)
    - \frac A{2}\sqrt{1+\bv\2}\,.
\end{equation}

In his paper~\cite[page~199]{matsyuk:Bopp}, Fritz Bopp asserted: ``Der
klassischen Bewegung \"uberlagert sich eine Zitterbewegung, die
durch die neuen Variabeln $\bv$ und $\bp'$ beschrieben wird. Sie
f\"uhrt zu spinartigen Effekten\dots''\nopagebreak
{\renewcommand{\thefootnote}{\fnsymbol{footnote}}
\footnote[1]{Upon the classical motion some vibrational one
superimposes itself that is described by the new variables $\bv$
and $\bp'$. It leads to the effects of spin type\dots} }

Now the Hamilton function on $T^3M$ may be obtained from(\ref{matsyuk:cal H}):
\begin{equation}\label{matsyuk:cal H final}
\mathcal H=\wp u+\frac {1}{2a}\|u\|^3\wp'\2-\frac A 2 \|u\|+1\,.
\end{equation}
Alternatively, one could get the same expression directly from
the assertion
\begin{equation}\label{matsyuk:cal H directly}
\mathcal H=\wp u + \wp'\dot u - \mathcal L + 1\,,
\end{equation}
assuming $\mathcal L$ be taken from~(\ref{matsyuk:Bopp}). In view of~(\ref{matsyuk:rank}),
one cannot resolve the Legendre transformation~(\ref{matsyuk:cal p}) in full,
but nevertheless it is possible to eliminate the variable $\dot u$ from~(\ref{matsyuk:cal H directly}).
First we calculate the momenta for~(\ref{matsyuk:Bopp})
\begin{gather*}
 \wp'=\dfrac{a}{\|u\|^5}\left[u\2\dot u-(u\cdot\dot u)u\right] \\
 \wp= \dfrac{Au}{2\|u\|} {}-a\left[\dfrac{\ddot u}{\|u\|^3}
 -3\,\dfrac{u\cdot\dot u}{\|u\|^5}\,\dot u
 -\dfrac{u\cdot\ddot u}{\|u\|^5}\,u+\dfrac{\dot u\2}{2\|u\|^5}\,u
 +\dfrac{5}{2}\dfrac{(u\dot u)^2}{\|u\|^7}\,u
\right].
\end{gather*}
In the next step we express all those quantities in~(\ref{matsyuk:cal H directly})
wherein the $\dot u$ enters, in terms of $\wp'$ and $u$ alone
\begin{equation}\label{matsyuk:}
\left\{
\begin{aligned}
\wp'\dot u&=\dfrac{\|u\|^3}{a}\,\wp'\2 \\
\mathcal L_r&=\dfrac{\|u\|^3}{2a^2}\,\wp'\2\,,
\end{aligned}
\right.
\end{equation}
and substitute into~(\ref{matsyuk:cal H directly}) to finally
achieve the  Hamilton function~(\ref{matsyuk:cal H final}).

The approach to building up the Hamilton function in present paper
differs from
that of H.~S.~P.~Gr\"asser.  I was inspired by his treatment of
general Lagrange function, quadratic in velocities, in the
framework of Finsler space, of which ours is a very
special case. But the physical model herein considered demands to include
also the free particle term $\mathcal L_e$.

It is not of much labour now
to calculate the fourth-order Euler-Poisson equation of the variational
problem with the Lagrange function~(\ref{matsyuk:Bopp}) from the starting
point of the Hamilton system~(\ref{matsyuk:H})
with the expression~(\ref{matsyuk:cal H final}) in hand.
For~(\ref{matsyuk:H}) we have:
\begin{equation*}
\left\{
\begin{aligned}
 \dfrac{dx}{d\tau}&=\lambda u\,,
\\
\dfrac{du}{d\tau}&=\lambda \dfrac{\|u\|^3}{a}\,\wp'+\mu u\,,
\\
\dfrac{d\wp}{d\tau}&=0\,, \\
\dfrac{d\wp'}{d\tau}&=\dfrac{A}{2}\lambda \dfrac{u}{\|u\|}
        -\lambda \wp-\dfrac{3}{2}\lambda \dfrac{\|u\|}{a}\,\wp'\2u-\mu\wp'\,.
\end{aligned}
\right.
\end{equation*}
The multiplier~$\mu$ may be obtained from the second equation by contracting it with~$u$ and recalling the Zermelo condition~(\ref{matsyuk:Z1}):
\begin{equation*}
    \mu=\frac{u\cdot u}{\|u\|^2}\,.
\end{equation*}
Only at this stage one has the right to put some constraints on the choice of
the parameter $\tau$. We put $u\2=1$ to obtain
\begin{subequations}\label{matsyuk:H Zitterbewegung u=1}
\renewcommand{\theequation}{\theparentequation.\arabic{equation}}
\begin{align}
\dfrac{du}{d\tau}&=\dfrac{\wp'}{a} \label{matsyuk:dot u u=1}\\
\dfrac{d\wp'}{d\tau}&=\dfrac{A}{2}\,u
        -\wp-\dfrac{3}{2a}\wp'\2 u\,, \label{matsyuk:.wp' u=1}
\end{align}
\end{subequations}
and it is clear that $\lambda=1$ and $\mu=0$, so that the
evolution equation~(\ref{matsyuk:Poisson}) regains the traditional shape now.

Next we differentiate the equation~(\ref{matsyuk:dot u u=1}) and substitute
the equation~(\ref{matsyuk:.wp' u=1}) therein, to obtain
\begin{gather}
\ddot u =   \dfrac{A}{2a}\,u
                     -\dfrac{\wp}{a}-\dfrac{3}{2a^2}\wp'\2 u\,,
 \label{matsyuk:..u}
\\
\dfrac{\wp\dot u}{a}={}- \ddot u \cdot \dot u \,,\label{matsyuk:..u.u}
\end{gather}
and, on the other hand, the contraction of (\ref{matsyuk:dot u u=1}) with
(\ref{matsyuk:.wp' u=1}) gives
\begin{equation}\label{matsyuk:.wp'wp' u=1}
\wp'\cdot \dot\wp'={}-a\,\wp\dot u\,.
\end{equation}
Differentiating (\ref{matsyuk:..u}) once again produces
\begin{equation}\label{matsyuk:...u}
\dddot u = \dfrac{A}{2a}\,\dot u{}-\dfrac{3}{a^2}\,(\wp'\cdot \dot\wp')\,u
        - \dfrac{3}{2a^2}\,\wp'\2\dot u\,,
\end{equation}
in where we substitute (\ref{matsyuk:.wp'wp' u=1}), (\ref{matsyuk:dot u u=1}),
 and, sequentially,
(\ref{matsyuk:..u.u}), to finish at the resulting fourth-order
equation of motion
\begin{equation}\label{matsyuk:...u final}
\dddot u + \left(\dfrac{3}{2}\,\dot u\2 - \dfrac{A}{2a}\right)\,\dot u
        +3\,(\dot u \cdot \ddot u)\,u=0\,.
\end{equation}
As soon as in the actual parametrization $k^2=\dot u\2$, on the constrained
manifold of constant relativistic acceleration $k_\so$
 the equation~(\ref{matsyuk:...u final}) {\em reduces to the equation of helical
 motion of relativistic spinning particle considered first by F.~Riewe~\cite{matsyuk:Riewe} and then by
G.~C.~Constantelos~\cite{matsyuk:Constantelos}:}
 \begin{equation}\label{matsyuk:Riewe}
\dfrac{d^2}{ds^2}\,\ddot x^\alpha +\varpi^2\ddot x^\alpha =0\,,
\end{equation}
where we have put
 $\varpi=\frac{3}{2}\,k_\so^2-\frac{A}{2a}$.

\bigskip
In the previous paper~\cite{matsyuk:Matsyuk}, I proved that {\em this fourth order
 equation of motion~(\ref{matsyuk:Riewe}) may be rigorously developed from
 the third order general equation of motion of classical dipole particle
 proposed by Mathisson in 1937 in~\cite{matsyuk:Mathisson}}.

\end{document}